\documentclass[journal]{IEEEtran}

\usepackage[utf8]{inputenc}
\usepackage[T1]{fontenc}
\usepackage{lmodern}
\usepackage{amsmath,amssymb}
\usepackage{graphicx}
\usepackage{hyperref}
\usepackage{cite}

\begin{document}

\title{Audio Palette: A Diffusion Transformer with Multi-Signal Conditioning for Controllable Foley Synthesis}

\author{Junnuo~Wang\\
New York University}

\markboth{}%
{Wang: Audio Palette: A Diffusion Transformer with Multi-Signal Conditioning for Controllable Foley Synthesis}

\maketitle

\begin{abstract}
Recent advances in diffusion-based generative models have enabled high-quality text-to-audio synthesis, but fine-grained acoustic control remains a significant challenge in open-source research. We present Audio Palette, a diffusion transformer (DiT) based model that extends the Stable Audio Open architecture to address this “control gap” in controllable audio generation. Unlike prior approaches that rely solely on semantic conditioning, Audio Palette introduces four time-varying control signals, loudness, pitch, spectral centroid, and timbre, for precise and interpretable manipulation of acoustic features. The model is efficiently adapted for the nuanced domain of Foley synthesis using Low-Rank Adaptation (LoRA) on a curated subset of AudioSet, requiring only 0.85\% of the original parameters to be trained. Experiments demonstrate that Audio Palette achieves fine-grained, interpretable control of sound attributes. Crucially, it accomplishes this novel controllability while maintaining high audio quality and strong semantic alignment to text prompts, with performance on standard metrics such as Fréchet Audio Distance (FAD) and LAION-CLAP scores remaining comparable to the original baseline model. We provide a scalable, modular pipeline for audio research, emphasizing sequence-based conditioning, memory efficiency, and a three-scale classifier-free guidance mechanism for nuanced inference-time control. This work establishes a robust foundation for controllable sound design and performative audio synthesis in open-source settings, enabling a more artist-centric workflow in the broader context of music and sound information retrieval.
\end{abstract}

\begin{IEEEkeywords}
text-to-audio synthesis, diffusion models, controllable generation, Foley sound, music information retrieval.
\end{IEEEkeywords}

\section{Introduction}

Generative models have made significant strides in image, video, and audio synthesis, with diffusion-based architectures becoming a state-of-the-art approach for high-fidelity generation [1], [2]. In audio research and MIR, diffusion models and latent diffusion models have enabled impressive results for text-to-audio (TTA) tasks, producing high-quality audio from natural language descriptions, in both autoregressive and diffusion-based systems [3], [4], [5]. Architectures such as Stable Audio Open build upon the Diffusion Transformer (DiT) to generate coherent, high-fidelity audio sequences from text prompts by operating in a continuous latent space [6], [7]. Recent work on latent TTA models, including AudioLDM and CLAP-based latent diffusion pipelines, has shown that contrastive language–audio representations can be leveraged for efficient and scalable text-to-audio generation [3], [8].

Despite these advances, a critical “control gap” persists. While TTA models excel at interpreting semantic content (for example, “a dog barking”), they largely fail to capture the performative aspects of sound: dynamic intensity, pitch contour, and textural evolution over time. This limitation is a bottleneck for professional applications such as film scoring, game audio design, and particularly Foley synthesis, where timing, nuance, and emotional weight are paramount [9], [10], [11], [12]. Traditional Foley artistry is an inherently gestural craft; a Foley artist performs the sound of a character’s footstep, conveying weight, emotion, and intent through subtle sonic variations [9]. This performative detail is essential for creating an immersive diegetic world, yet purely text-driven systems rarely provide a mechanism to specify such nuances.

Moreover, while several proprietary systems and recent research prototypes offer advanced control or video-guided Foley generation, the open-source ecosystem remains limited in frameworks that combine multi-modal conditioning in a unified manner [3], [7], [8], [12]. This scarcity constrains research into expressive and interactive synthesis paradigms, including those directly relevant to MIR, such as controllable sound effects for interactive music systems, audio–visual alignment, and gesture-driven sound design [5], [13]. At the same time, traditional Foley workflows are labor-intensive, rely on specialized spaces and props, and do not scale easily to large or interactive media catalogs.

To address these challenges, we propose Audio Palette, a DiT-based model that extends Stable Audio Open to enable fine-grained, interpretable control over sound generation in a Foley context. This work makes the following contributions:

First, we introduce a multi-signal conditioning framework that augments a state-of-the-art open-source TTA model with four distinct time-varying acoustic control signals (loudness, pitch, spectral centroid, and timbre), enabling precise and reproducible synthesis guided by both semantic and acoustic specifications [7], [14], [15]. This transforms the generative process into a performative act, aligning it more closely with the craft of Foley artistry [9]. Second, we demonstrate an efficient, specialized Foley synthesizer by applying LoRA-based parameter-efficient fine-tuning on a curated subset of AudioSet, making it feasible to adapt a large model to a nuanced Foley domain without retraining all parameters [16], [17]. Third, we propose a multi-scale classifier-free guidance mechanism that splits guidance into semantic, dynamic, and timbral components, offering disentangled user control at inference time and aligning with related work on prosody and style transfer in speech [18], [19].

\section{Related Work}

\subsection{Text-to-Audio Synthesis with Diffusion Models}

Denoising Diffusion Probabilistic Models (DDPMs) and score-based models have become central tools for high-fidelity audio generation, following their success in image synthesis [1], [20]. Early audio diffusion models such as DiffWave and WaveGrad focused on waveform-domain generation conditioned on mel-spectrograms, demonstrating that non-autoregressive diffusion can match or surpass autoregressive vocoders in quality while being more parallel and scalable [20], [21].

Recent work has shifted toward latent diffusion for audio, paralleling developments in image generation. AudioLDM, for instance, trains a latent diffusion model in the space of CLAP-based audio embeddings while conditioning on text, enabling efficient text-to-audio, style transfer, and zero-shot manipulations [3], [4]. At the same time, contrastive language–audio pretraining (CLAP) has established large-scale audio–text representation learning as a key building block for TTA [8].

Stable Audio Open extends this line of work by providing an open-weights text-to-audio model based on a DiT backbone operating on continuous latent representations [7]. It supports stereo 44.1 kHz audio, transparent data documentation, and competitive performance compared to closed-source baselines, making it a natural foundation for controllable extensions [7]. Audio Palette builds directly on this DiT-based latent diffusion framework and focuses on introducing explicit, interpretable control signals for Foley synthesis.

\subsection{Controllable Audio Generation}

Controllable audio generation aims to balance new conditioning mechanisms with preservation of audio quality and semantic coherence. In TTS, prosody control via pitch, duration, and energy predictors has become standard; FastPitch, for example, conditions mel-spectrogram generation on predicted F0 contours, which can be manipulated to modulate perceived expression [18]. Tacotron extensions for prosody transfer learn a latent prosody embedding from reference audio and use it to control the style and rhythm of synthesized speech, providing a reference-based interface for expressive synthesis [19]. These works illustrate the value of time-varying and reference-based control in generative speech, which we adapt to general sound and Foley.

For non-speech audio, several lines of research have explored controllable synthesis. Diffusion-based Foley models for DCASE tasks condition on event labels and temporal event features to ensure synchronization to on-screen actions [10], [11], [22]. Video-guided Foley generation further incorporates visual motion and scene dynamics as conditioning signals [12]. In parallel, CLAP-based systems and AudioLDM explore text-conditioned style transfer, yet generally do not expose low-level acoustic controls [3], [8].

Closer to our work, Sketch2Sound proposes a controllable TTA DiT that conditions on loudness, brightness, and pitch control signals extracted from vocal or sonic imitations, combined with text prompts [14]. It introduces random median filtering of control signals to support sketch-like, temporally imprecise gestures, and demonstrates that a lightweight conditioning head with modest fine-tuning suffices to add powerful controls on top of a pre-trained DiT [14], [15]. Audio Palette is conceptually aligned with Sketch2Sound but makes two key contributions: it introduces an explicit timbre trajectory via MFCCs and explores a three-scale classifier-free guidance scheme that disentangles semantic, dynamic, and timbral adherence during sampling.

Within MIR, controllable and structured generation has been widely studied for symbolic music and audio, for example, through chord-conditioned melody generation, timbre-controlled synthesis, and structured latent representations [23], [24], [25], [26]. Our work complements these efforts by targeting Foley sound rather than music and by focusing on interpretable continuous control in a DiT-based latent diffusion model.

\section{Methodology}

\begin{figure}[t]
\centering
\includegraphics[width=\linewidth]{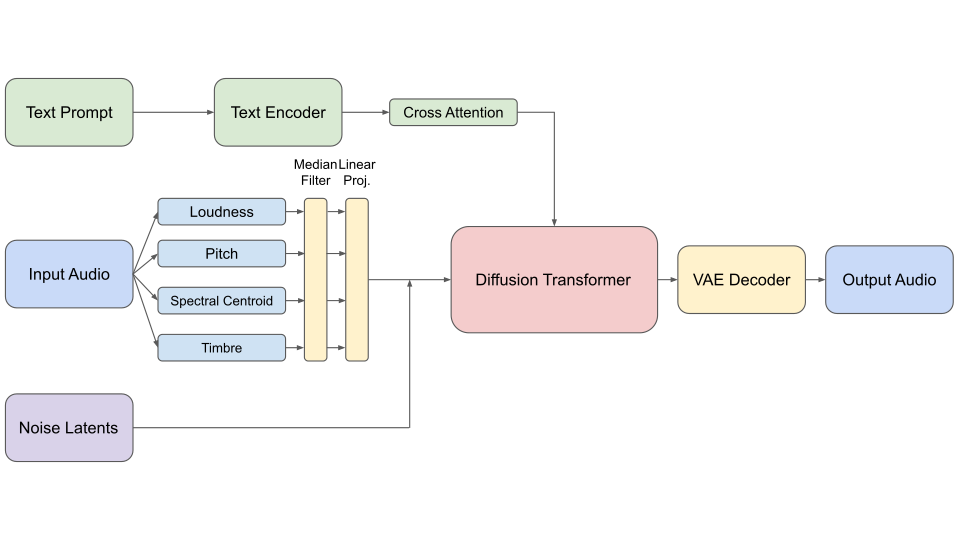}
\caption{Overview of the Audio Palette architecture.}
\label{fig:audio_palette_overview}
\end{figure}

Figure~\ref{fig:audio_palette_overview} provides an overview of the Audio Palette architecture, which builds upon Stable Audio Open and introduces a multi-signal conditioning pathway and parameter-efficient fine-tuning.

\subsection{Architectural Foundation}

The base model follows the Stable Audio Open architecture. It consists of three main components: A variational autoencoder (VAE) encodes stereo audio at 44.1 kHz into a compressed latent representation at a low frame rate, reducing sequence length and enabling efficient latent diffusion [7]. A corresponding decoder reconstructs the waveform from denoised latents. The VAE bottleneck has latent channels of size 64. A pre-trained, frozen T5-based text encoder produces semantic embeddings from natural language prompts, in line with earlier TTA and text-to-image diffusion systems. The DiT backbone operates as the core generative model: given a noisy latent tensor $z_t$ at timestep $t$, it predicts the added noise $\varepsilon$ and is trained using a standard diffusion loss [6]. Semantic text embeddings are injected as conditioning via cross-attention within transformer blocks, similar to other DiT-based latent diffusion models [3].

\subsection{Multi-Signal Conditioning Module}

The primary architectural contribution of Audio Palette is the integration of four time-varying control signals, which guide the diffusion process alongside the text prompt.

For a given reference audio, we extract four trajectories that capture complementary aspects of the sound. Loudness is computed as a per-frame RMS energy contour. Pitch is estimated using CREPE, a deep convolutional pitch tracker that operates directly on the waveform and achieves state-of-the-art monophonic pitch estimation performance [15]. Spectral centroid is computed as the energy-weighted mean frequency per frame, serving as a proxy for perceptual brightness. Timbre is captured via the first 13 MFCCs, a standard representation of spectral shape and widely used in MIR and audio classification [13], [27], [28].

All signals are computed at a frame rate that is aligned (or resampled) to the VAE latent sequence using linear interpolation. The loudness, pitch, and spectral centroid are concatenated into a “dynamic control” tensor, while the MFCC stack forms a “timbre control” tensor. Both are concatenated along the feature axis to form a single control tensor, which is passed through a trainable linear projection to match the latent channel dimension of the DiT. This low-capacity conditioning head is inspired by Sketch2Sound’s approach of using lightweight linear layers for control injection [14].

The projected control embeddings are added element-wise to the noisy latents $z_t$ at each denoising step, at the appropriate temporal resolution. This direct fusion approach modifies the latent trajectory with minimal architectural change and maintains compatibility with the original DiT design. Similar latent-space conditioning has been shown to be effective for controllable synthesis in both image and audio diffusion models [6], [14], [20], [21].

\subsection{Parameter-Efficient Fine-Tuning}

To adapt Stable Audio Open to a specialized Foley synthesis domain, we use parameter-efficient fine-tuning via LoRA [17]. The goal is to modify the DiT’s behavior while leaving the majority of weights frozen.

We curate a $\sim$150-hour dataset from AudioSet, a large-scale collection of 10-second YouTube clips annotated with sound event labels [16]. We select classes that correspond to common Foley categories such as footsteps, impacts, animal sounds, human actions, and environmental noises, guided by the AudioSet ontology and by classes used in DCASE 2023 Foley synthesis tasks [22]. We retrieve text descriptions based on class labels, tag metadata, and heuristic templates, following similar practices in TTA work [3], [7].

Fine-tuning proceeds in a self-supervised fashion. For each audio–text pair, we encode the audio with the VAE to obtain a latent sequence, add forward diffusion noise, and train the DiT to predict the noise while being conditioned on the text embeddings and the four extracted control trajectories. The target is to reconstruct the original audio’s latent trajectory; thus, the model is trained to interpret the control signals as a specification of dynamics and timbre for that sound class.

LoRA is applied to the query and value projection matrices in the attention layers of the DiT, inserting low-rank adapters while keeping the original weights frozen. LoRA has been shown to drastically reduce trainable parameters and memory while preserving performance across NLP and vision models [17]. In our configuration, the LoRA rank is 16, leading to approximately 0.85\% of the DiT parameters being trained. The T5 encoder and the VAE remain frozen. We train the control projection layers and LoRA weights using the AdamW optimizer for 40{,}000 steps.

To enhance robustness, we apply independent dropout to each control signal embedding and to the text embedding. This encourages the network not to rely excessively on any single conditioning source. In addition, we apply random median filtering to the control trajectories, following Sketch2Sound’s insight that smoothing controls during training improves generalization to sketch-like inputs [14]. During training, each control trajectory is randomly either left untouched or passed through a median filter with a window chosen from a small set of values. This encourages the model to focus on the coarse contour of dynamics and timbre rather than overfitting to high-frequency artifacts. A similar strategy of smoothing trajectories has been useful in video and event-guided Foley synthesis [10], [12], [22].

\subsection{Multi-Scale Classifier-Free Guidance}

Classifier-free guidance (CFG) is widely used in diffusion models to trade off sample diversity and adherence to conditioning [29]. In its standard form, a single guidance scale is applied to the difference between conditional and unconditional predictions.

Audio Palette extends CFG to three scales, reflecting the three distinct conditioning pathways: semantic text, dynamic controls, and timbral controls. Specifically, we train conditional variants where subsets of the conditioning inputs are masked, allowing us to estimate conditional and partially conditional scores during inference. We define guidance scales $s_{\text{text}}$, $s_{\text{ctrls}}$, and $s_{\text{timbre}}$, which respectively weight the contribution of text, dynamic controls (loudness, pitch, spectral centroid), and timbral controls (MFCCs).

At inference time, users can set $s_{\text{text}}$ to emphasize semantic correctness of the generated sound relative to the prompt, $s_{\text{ctrls}}$ to enforce tight adherence to dynamic trajectories, and $s_{\text{timbre}}$ to prioritize timbre transfer from the reference audio. This multi-scale CFG mechanism can be seen as an analogue to separate control knobs for prosody and style in expressive TTS [18], [19]. It provides a flexible artist-in-the-loop interface, allowing users to balance fidelity to text, gesture-like dynamics, and reference timbre.

\section{Experiments and Results}

We evaluate Audio Palette on audio quality, semantic alignment, and controllability. All experiments are conducted on a held-out test split from the curated AudioSet Foley subset.

\subsection{Experimental Setup}

The test set consists of AudioSet clips from Foley-relevant classes not used during fine-tuning, ensuring that the model is evaluated on unseen examples but within the same domain [3], [16]. For each clip, we use its label-derived text description as the prompt and extract control signals from the clip itself. The baseline is the original, unmodified Stable Audio Open model, which is text-only and has no access to control trajectories [7].

We focus on two widely used objective metrics. Fréchet Audio Distance (FAD) measures the distance between Gaussian distributions fitted to embeddings of real and generated audio, using a pre-trained VGGish encoder [30]. Lower scores indicate closer match to real audio and higher overall quality. LAION-CLAP similarity is computed as the cosine similarity between CLAP embeddings of the generated audio and the corresponding text prompt [8]. Higher scores indicate stronger semantic alignment and are common in TTA evaluations.

Audio Palette and the baseline generate one 10-second clip per test prompt. For the baseline, only the text prompt is provided. For Audio Palette, both the prompt and the corresponding reference control trajectories (extracted from the original clip) are used. Unless otherwise noted, guidance scales are set to default values calibrated on a small validation set, and generation time and computational budgets are matched across models.

\subsection{Quantitative Analysis}

\begin{table}[t]
\centering
\caption{Main Quantitative Results on the Foley Test Set}
\label{tab:main_results}
\begin{tabular}{lcc}
\hline
Model & FAD (↓) & CLAP Score (↑) \\
\hline
Stable Audio Open 1.0 (Baseline) & 5.82 & 0.615 \\
Audio Palette (Ours)             & 5.95 & 0.589 \\
\hline
\end{tabular}
\end{table}

Table~\ref{tab:main_results} reports FAD and CLAP scores on the Foley test set. Audio Palette exhibits a small increase in FAD and a modest decrease in CLAP score relative to the text-only baseline. This pattern is consistent with prior work, where additional conditioning (for example, control signals or temporal event features) often introduces a slight cost in distributional quality and semantic alignment in exchange for more constrained and controllable generation [3], [10], [14]. The key observation is that the degradation is modest and remains within a range comparable to differences between strong baselines in TTA benchmarks.

Given that Audio Palette must satisfy both text prompts and precise time-varying control trajectories, these results suggest that the DiT backbone and LoRA adapters are capable of integrating multi-signal conditioning without severely compromising audio fidelity or text alignment.
\subsection{Qualitative Analysis}

Objective metrics such as FAD and CLAP do not directly measure controllability. To assess whether the model follows control trajectories in a perceptually meaningful way, we conduct qualitative generation studies.

For loudness, we use a prompt such as “A dog barking, starting quiet, getting loud, then quiet again” and an accompanying vocal imitation that follows a crescendo–decrescendo pattern. The generated dog barks track the rising and falling RMS envelope of the reference, yielding a clear correspondence between control contour and perceived intensity. Similar examples are widely used in Sketch2Sound and TTS prosody control work [14].

For pitch, we prompt “A siren with a rising pitch” and provide an ascending sine wave as the reference. The generated siren exhibits a continuous upward pitch glide aligned with the reference F0 contour. For spectral centroid, we use “A cymbal crash that fades out” combined with filtered noise whose low-pass cutoff decreases over time; the generated sound is initially bright and gradually darkens, tracking the centroid trajectory.

Timbre transfer is assessed with prompts like “Footsteps on gravel” accompanied by a reference audio of crunching leaves. Audio Palette generates footstep-like events with the temporal pattern of walking but with a brittle, leafy timbre, illustrating the ability to decouple event structure from timbral character. This mirrors prosody transfer in TTS, where prosodic style is transferred independently of text [19] and is conceptually related to style transfer in AudioLDM [3].

The multi-scale guidance mechanism enhances this flexibility. Increasing $s_{\text{timbre}}$ yields outputs that more closely match the reference MFCC trajectory, often at the cost of minor deviations in dynamics or prompt specificity. Increasing $s_{\text{ctrls}}$ enforces more precise adherence to the dynamic contour, even if the model has to compromise slightly on exact semantic nuances of the text prompt, consistent with the typical CFG trade-off [6], [20], [21]. These behaviors align with qualitative reports in prior controllable diffusion work for audio and video-guided Foley [10], [12], [14], [22].

\subsection{Ablation Studies}

To quantify the impact of different control subsets, we train several ablation variants and evaluate them with FAD and CLAP, as summarized in Table~\ref{tab:ablation}.

\begin{table}[t]
\centering
\caption{Ablation Study on Control Signals}
\label{tab:ablation}
\begin{tabular}{lcc}
\hline
Model Configuration        & FAD (↓) & CLAP Score (↑) \\
\hline
Baseline (Text Only)       & 5.82 & 0.615 \\
+ Loudness, Pitch, Centroid & 5.98 & 0.595 \\
+ Timbre (MFCCs) only      & 5.90 & 0.605 \\
Full Model (All Signals)   & 5.95 & 0.589 \\
\hline
\end{tabular}
\end{table}

Introducing loudness, pitch, and spectral centroid together yields the largest decrease in CLAP, which is expected given that these dynamic controls impose strong temporal structure that may occasionally conflict with the text prior. Timbre-only conditioning leads to a smaller shift in both metrics, suggesting that specifying a broad spectral shape is a relatively weak constraint. The full model, combining all four signals, finds a middle ground between the two configurations, preserving much of the controllability while mitigating the cumulative impact on FAD.

This pattern mirrors observations in earlier controlled synthesis work, where tighter constraints on temporal structure can lead to subtle trade-offs in perceived naturalness or semantic “freedom” [14]. The ablations also confirm that the MFCC-based timbre trajectory provides an additional control channel that can be used in isolation or combined with dynamics, without overwhelming the baseline DiT.

\section{Conclusion}

We have presented Audio Palette, a DiT-based latent diffusion model for controllable audio generation, specialized for Foley synthesis. Built upon the Stable Audio Open architecture, Audio Palette introduces four time-varying acoustic control signals, loudness, pitch, spectral centroid, and timbre. It uses LoRA-based fine-tuning to adapt a general-purpose text-to-audio model to a Foley domain with minimal parameter updates [7], [16], [17]. Our experiments show that Audio Palette offers precise, interpretable control over these acoustic dimensions while maintaining high perceptual quality and competitive text–audio semantic alignment, as measured by FAD and LAION-CLAP[8], [30]. The proposed multi-scale classifier-free guidance mechanism provides an intuitive interface for balancing semantic, dynamic, and timbral adherence during inference.

At the same time, the system has several limitations. It relies on reference audio for control extraction and cannot yet infer plausible control trajectories directly from text alone. Very aggressive guidance scales can introduce artifacts, requiring users to tune settings per prompt and control configuration, similar to other CFG-based diffusion models [6]. Moreover, the model is fine-tuned specifically for Foley-like content and is not evaluated on complex musical mixtures or dense ambient soundscapes; additional adaptation or domain-specific training would be needed for such tasks [3], [13].

These limitations suggest several directions for future work. One avenue is the integration of Audio Palette into interactive interfaces that allow sketching, editing, and sequencing of control contours, lowering the barrier for non-expert creators and aligning with gesture-based MIR tools. Another is to incorporate visual conditioning from video, using motion or scene dynamics to infer control trajectories, building on recent work in video-guided Foley [12]. Finally, extending the framework to broader MIR benchmarks and creative tools would help clarify how controllable diffusion-based sound models can be integrated into music production, interactive media, and sound design ecosystems.

\end{document}